\documentclass[reprint,aps,prb,english,superscriptaddress]{revtex4-2}

\usepackage{graphicx}
\usepackage{bm,amssymb,amsmath,mathrsfs,amstext,latexsym}

\usepackage{physics}
\usepackage{enumitem}
\usepackage{graphicx}

\usepackage[usenames,dvipsnames]{xcolor}
\usepackage[colorlinks=true,citecolor=blue,urlcolor=blue,linkcolor=blue]{hyperref}
\usepackage[normalem]{ulem}
\usepackage{booktabs}

\let\tilde\relax
\newcommand{\tilde}[1]{\widetilde{#1}}

\begin{document}

\title{Effective delocalization in the one-dimensional Anderson model with stealthy disorder}

\author{Carlo Vanoni}
\affiliation{Department of Physics, Princeton University, Princeton, New Jersey, 08544, USA}

\author{Jonas Karcher}
\affiliation{Max Planck Institute for the Physics of Complex Systems, N\"othnitzer Str. 38, 01187 Dresden, Germany}

\author{Mikael C. Rechtsman}
\affiliation{Department of Physics, The Pennsylvania State University, University Park, PA, USA}

\author{Boris L. Altshuler}
\affiliation{Physics Department, Columbia University, 538 West 120th Street, New York, New York 10027, USA}

\author{Paul J. Steinhardt}
\email{steinh@princeton.edu}
\affiliation{Department of Physics, Princeton University, Princeton, New Jersey, 08544, USA}

\author{Salvatore Torquato}
\affiliation{Department of Chemistry, Princeton University, Princeton, New Jersey 08544, USA}
\affiliation{Department of Physics, Princeton University, Princeton, New Jersey, 08544, USA}
\affiliation{Princeton Institute for the Science and Technology of Materials, Princeton University, Princeton, New Jersey 08544, USA}
\affiliation{Program in Applied and Computational Mathematics, Princeton University, Princeton, New Jersey 08544, USA}

\date{\today}


\begin{abstract}
    We study analytically and numerically the Anderson model in one dimension with “stealthy" disorder, defined as having a power spectrum that vanishes in a continuous band of wave numbers. Motivated by recent studies on the optical transparency properties of stealthy hyperuniform layered media, we compute the localization length using a perturbative expansion of the self-energy. We find that, for fixed energy and small but finite disorder strength $W$, there exists for any finite length system a range of stealthiness $\chi$ for which the localization length exceeds the system size. This kind of “effective delocalization" is the result of the novel kind of correlated disorder that spans a continuous range of length scales, a defining characteristic of stealthy systems. Unlike uncorrelated disorder, for which the localization length $\xi$ scales as $W^{-2}$ to leading order for small W, the leading order terms in the perturbation expansion of $\xi$ for stealthy disordered systems vanish identically for a progressively large number of terms as $\chi$ increases such that $\xi$ scales as $W^{-2n}$ with arbitrarily large $n$. Moreover, we support our analytical results with numerical simulations. Our results introduce stealthy disorder into quantum tight-binding models and show that enforcing a low-$k$ spectral gap markedly alters the scattering landscape, enabling localization lengths that exceed the system size at fixed disorder strength. Since this mechanism relies only on the spectral properties of the disorder, it carries over directly to photonic and phononic wave systems.
\end{abstract}

\maketitle

\section{Introduction}

Since the seminal work by P. W. Anderson in the late 1950s~\cite{Anderson1958absence}, the interplay between randomness and quantum effects in non-interacting systems has received considerable attention in the condensed matter community~\cite{evers2008anderson}. The impact of space dimensionality has been addressed in the scaling theory of localization~\cite{abrahams1979scaling} and in recent investigations~\cite{ueoka2014dimensional,Tarquini2017critical,altshuler2025Renormalization}, including the infinite-dimensional limit~\cite{tikhonov2016Anderson,bera2018return,sierant2023universality,vanoni2023renormalization,Derrida1980REM,Baldwin2016ManyBody,Balducci2025Scaling}, relevant for the interacting case, known as many-body localization~\cite{altshuler1997quasiparticle,Basko06,oganesyan2007localization,ros2015integrals,tikhonov2021AndersonMBL,niedda2024renormalizationgroupanalysismanybodylocalization}. 
In one and two dimensions and for a random uncorrelated onsite potential, it is well known~\cite{abrahams1979scaling} that any small amount of disorder (whose strength $W^2$ is parametrized by its variance) leads to the localization of the wave functions. In 1D, the localization length $\xi$ at small disorder is proportional to $W^{-2}$~\cite{Lee1985Disordered}.

When introducing correlations in the potential energy, the scenario can change and show interesting new phenomena. This occurs, for instance, in the case of quasi-periodic systems~\cite{Aubry1980Annals,Luschen2018}, displaying extended, critical, and localized phases, separated by non-trivial mobility edges~\cite{Ganeshan2015Nearest,Biddle2010Predicted,Goncalves2023Critical}. 
Another example, relevant to polymer physics, is that of random dimers~\cite{Dunlap1990Absence,Phillips1991Localization}, where a particle can become delocalized in one dimension if two uncorrelated random energies are assigned at random to pairs of lattice sites. However, only two states are extended in this case.
A further interesting case that has received attention in the literature, also because of its experimental relevance in cold-atom ~\cite{Sanchez2007Anderson,Billy2008} and photonics \cite{schwartz2007transport,dikopoltsev2022observation} experiments, is that of speckle potentials, in which the two-point correlation function has a high-momentum cutoff. 
In one-dimensional (1D) systems, speckle correlations suppress back-scattering and can lead to an effective suppression of Anderson localization~\cite{Lugan2009Onedimensional,Piraud2012Anderson,morpurgo2025localizationtransitioninteractingquantum}.
The effect of correlated disorder on Anderson localization has also been studied previously in a series of papers~\cite{izrailev_localization_1999,izrailev_anomalous_2012,Izrailev1995Hamiltonian}, where it has been shown that the inverse localization length is proportional to the product of the Fourier transform of the correlation of the disorder and $W^2$~\cite{izrailev_localization_1999}. 
According to these results, the localization length should diverge if the power spectrum vanishes. 
However, these results should be regarded as a perturbative outcome including only the first-order term at small $W$. Higher-order terms thus lead to a finite localization length.

A special type of disorder is provided by hyperuniform systems~\cite{To03a,To18a}, which are characterized by an anomalous suppression of large-scale density fluctuations. In momentum space, they are characterized by a structure factor $\mathcal{S}(k)$ that vanishes as $k \to 0$. Since the first studies in the early 2000s, hyperuniform systems have appeared in a wide variety of systems, ranging from number theory~\cite{To18d,To08b} to biological systems~\cite{jiao_avian_2014,Ma15} (see Ref.~\cite{To18a} for a review).

Among hyperuniform disordered systems, there is a special class that is referred to as stealthy hyperuniformity~\cite{uche_constraints_2004,batten_classical_2008,Zh15a,Zh15b}. In this context, stealthy means that the structure factor of such systems vanishes for a continuous range of wave numbers. Therefore, a stealthy hyperuniform system is such that 
\begin{equation*}
    \mathcal{S}(k)=0 \quad \mathrm{for} \ 0 \leq k<k_0.
\end{equation*}

It is useful to introduce the stealthiness parameter $0\leq \chi \leq 1/2$, defined as the relative fraction of constrained wavevectors. For point patterns, the configuration converges to a periodic lattice in the limit $\chi \to 1/2$. For the specific case 1D systems, relevant in this paper, $\chi = k_0/(2\pi)$~\cite{batten_classical_2008}.
Stealthy hyperuniform systems display remarkable properties, including superior transport~\cite{To21d}, elasticity~\cite{kim_effective_2020}, and wave propagation properties~\cite{kim_effective_2023,Vanoni2025Dynamical} compared to non-stealthy and non-hyperuniform isotropic amorphous states of matter~\cite{To18a}.

In a recent study, Klatt et al.~\cite{klatt2025transparencyversusandersonlocalization} studied the electromagnetic wave transmission in disordered stealthy hyperuniform layered media formed by slabs of alternating high and low dielectric constant. 
By computing the Lyapunov exponent using transfer matrix methods, they found no evidence of localization for system sizes up to 10000 slabs, similar to what is found in periodic systems. 
On the analytical side, it has been shown in Ref.~\cite{kim_effective_2023} that the nonlocal strong contrast expansion~\cite{torquato_nonlocal_2021} truncated at third-order predicts that 1D disordered stealthy hyperuniform layered media have perfect transparency (no Anderson localization) for a range of frequencies. Therefore, Anderson localization could only arise at fourth order and beyond.
It should be mentioned that the nonlocal strong contrast expansion is characterized by fast convergence, and lower-order terms provide accurate approximations for the effective dielectric constant.

Motivated by the aforementioned effects of correlated disorder on Anderson localization and the recent results on the effect of stealthy hyperuniformity on 1D layered media, in this work, we study the effect of stealthy disorder in the 1D Anderson model.
In this context, the stealthiness parameter $\chi$ measures the fraction of harmonics for which the power spectrum of the random potential vanishes (see Eq.~\eqref{eq:stealthiness}). Unlike the quasi-periodic models, in this case, the system is disordered and allows a continuous range of energies, unlike the random dimer model.
It should be mentioned that the effect on localization of non-stealthy hyperuniform disorder in quantum systems has been addressed in the literature~\cite{Crowley2019Quantum,Shi2022Manybody}; a recent finding, for example, showed that Lifshitz tails have different scalings compared to uncorrelated disorder, effectively widening the band gap~\cite{Karcher2024Effect}. Experiments on disordered two-dimensional stealthy photonic crystal systems have demonstrated non-trivial wave transport properties with both strong disorder and non-Hermiticity \cite{barsukova2025stealthy}. However, the latter work did not examine localization behavior.

We determine the localization length using a perturbation theory expansion of the self-energy at small disorder $W \ll 1$ for the special case of stealthy hyperuniform disorder. Our procedure can be applied to any kind of stealthy disorder.
We are able to determine, for each energy and size of the stealthy region, the power of $W$ responsible for the leading contribution to the localization length. In particular, we find that, for any given length of the 1D system $L$, energy and weak disorder $W \ll 1$, there exists a value of the stealthiness parameter $\chi$ in which the system is disordered but with localization length exceeding the system size, and thus effectively delocalized. If the thermodynamic limit $L \to \infty$ is taken first, the localization length is always finite, thus not violating localization theorems~\cite{Wegner1976,abrahams1979scaling}.

We support our analytical findings with a numerical diagonalization computation of the localization length obtained from the fractal dimension of eigenstates in systems of sizes up to $L=O(10^6)$. We find that the perturbative calculation predicts the correct scaling of the localization length when varying the size of the stealthy region, and locates correctly the transition point between regions with different scalings.

\section{Model}

We consider the Anderson model in one spatial dimension with the Hamiltonian $H = H_0 + V$ obtained by summing the tight-binding Hamiltonian $H_0 = -\sum_j (c_{j+1}^{\dagger} c_j + c_{j}^{\dagger} c_{j+1})$ and potential energy $V = \sum_j w_j c^{\dagger}_j c_j$, where $c_j^{\dagger}$ and $c_j$ are the creation and annihilation operators (respectively) for a particle at site $j$ of the chain, with periodic boundary conditions. The disorder is contained in the random numbers $w_j$, which will be specified later. We have set the hopping amplitude to unity, so that the strength of the disorder is the only relevant energy scale in the problem.

By introducing the creation and annihilation operators ($\tilde{c}_k^{\dagger}$ and $\tilde{c}_k$ respectively) for a particle of momentum $k$ through the relation $c_j = \frac{1}{\sqrt{L}} \sum_k e^{ikj} \tilde{c}_k$, we can write
\begin{gather}
     H_0 = \sum_k \epsilon_k \tilde{c}^{\dagger}_k \tilde{c}_k, \quad \epsilon_k = -2 \cos k, \label{eq:H0_Fourier} \\
     V = \frac{1}{L} \sum_{k,k'} \left( \sum_q \tilde{w}_q \delta_{k'-k,q} \right) \tilde{c}^{\dagger}_{k'} \tilde{c}_k = \sum_{k,q} \tilde{w}_q \tilde{c}^{\dagger}_{k+q} \tilde{c}_k, \label{eq:V_Fourier}
\end{gather}
where we used $\tilde{w}_q = \sum_j e^{-iqj} w_j$. In the following, we will always use the Fourier-space representation of $H_0$ and $V$ defined above; henceforth, we will drop the tildes to simplify the notation. 

We are interested in extracting the random numbers $w_j$, $j=1,\dots, L$, for a prescribed power spectrum
\begin{equation}
    S(q) = \frac{|w_q|^2}{W^2},
\end{equation}
and do so by means of a Fourier filtering procedure.
As a first step, we define the amplitudes
\begin{equation}
    A(q) = \sqrt{S(q)},
\end{equation}
with $S(q) = S(-q)$. We then extract complex Gaussian random variables $\eta_q$, i.e., distributed according to $p(\eta_q) = \mathcal{N}(0,1)$, so that $\langle \eta_q \eta_{q'}^* \rangle = \delta(q-q')$.
The Fourier components of the desired random variables are then given by
\begin{equation}
    w_q = W A(q) \, \eta_q
\end{equation}
with the constraint $w_q = w_{-q}^*$. In the previous expression, we have introduced the disorder strength $W$.
Following this procedure, we get random numbers $w_q$ such that
\begin{equation}
\label{eq:corrdis}
    \langle w_q \rangle = 0, \quad \langle w_q w_{q'} \rangle =  W^2 S(q) \, \delta(q+q').
\end{equation}
By construction, the higher-order moments $\langle w_{q_1} w_{q_2} \dots w_{q_n} \rangle$ are expressed in terms of the two-point function only for even values of $n$ and vanish for odd values of $n$.

\section{Perturbation theory}

\begin{figure}
    \centering
    \includegraphics[width=\linewidth]{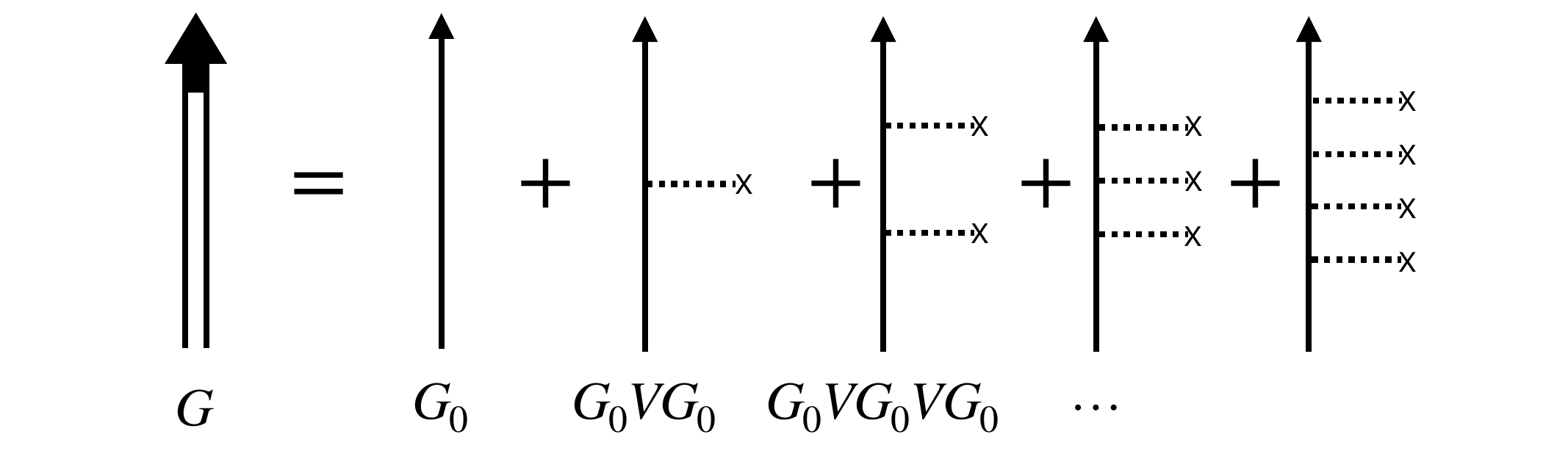}
    \caption{Graphical representation of Dyson equation, see Eq.~\eqref{eq:DysonEq} in the main text.}
    \label{fig:dysonEq}
\end{figure}

We are now interested in calculating the dependence of the localization length $\xi(k)$ on $W$ at a given momentum (or, equivalently, energy through the dispersion relation in Eq.~\eqref{eq:H0_Fourier}).
In 1D systems, the localization length is proportional to the mean free path $\ell(k)$, according to $\xi(k) \simeq 2 \ell(k)$~\cite{Lee1985Disordered}, which in turn is proportional to the mean survival time $\tau(k)$, defined as $\ell(k) = (\partial_k \epsilon_k) \tau(k)$~\footnote{The mean survival time $\tau$ can be interpreted as the average time between two scattering events of the quantum particle on the disordered potential}.
At this point, it is well known~\cite{Fetter2003Quantum} that the mean survival time is related to the imaginary part of the on-shell self-energy $\Sigma(k, \epsilon_k)$ according to
\begin{equation}
\label{eq:survtime}
    \frac{1}{\tau(k)} = -2 \, \mathrm{Im} \Sigma(k,\epsilon_k).
\end{equation}
In the weak disorder limit $W \ll 1$ (corresponding to the weak scattering limit $(k \ell)^{-1} \ll 1$), we can treat the potential $V$ as a small perturbation and compute the self-energy from the perturbation theory of the propagator
\begin{equation}
    G(k,E) = \frac{1}{E-\epsilon_k -V + i \varepsilon^+}.
\end{equation}
Following the standard textbook approach~\cite{Fetter2003Quantum,Akkermans_Montambaux_2007}, we can expand the propagator in powers of $V$ as
\begin{equation}
\label{eq:DysonEq}
    G = G_0 + G_0 V G_0 + G_0 V G_0 V G_0 + \dots
\end{equation}
where the bare propagator $G_0$ is given by
\begin{equation}
    G_0(k,E) = \frac{1}{E-\epsilon_k +i\varepsilon^+}.
\end{equation}
The Dyson equation~(\ref{eq:DysonEq}) can be represented graphically, as shown in Fig.~\ref{fig:dysonEq}.

After disorder averaging, only terms containing an even number of $V$ insertions will contribute, in virtue of Wick's theorem for Gaussian random variables. In addition, from Eq.~\eqref{eq:V_Fourier} we can see that the vertex of Feynman diagrams, given by $V$, has the effect of increasing the momentum $k \to k+q$, with momentum conservation at each vertex (see Eq.~\eqref{eq:corrdis}). 
With these observations, the propagator can be written as
\begin{equation}
    G = G_0 + G_0 \Sigma G,
\end{equation}
where $\Sigma(k, E)$ is the self-energy, i.e., it contains all diagrams without external $G_0$ legs and such that they cannot be split in two by removing a single $G_0$ line~\footnote{Sometimes in the many-body quantum physics literature $\Sigma(k, E)$ is referred to as the \emph{proper} self-energy, to distinguish it from the self-energy diagrams including contributions that are one-particle reducible. We will not make this distinction here, as all the relations we will need only include the proper self-energy.}.
\begin{figure}
    \centering
    \includegraphics[width=0.3\linewidth]{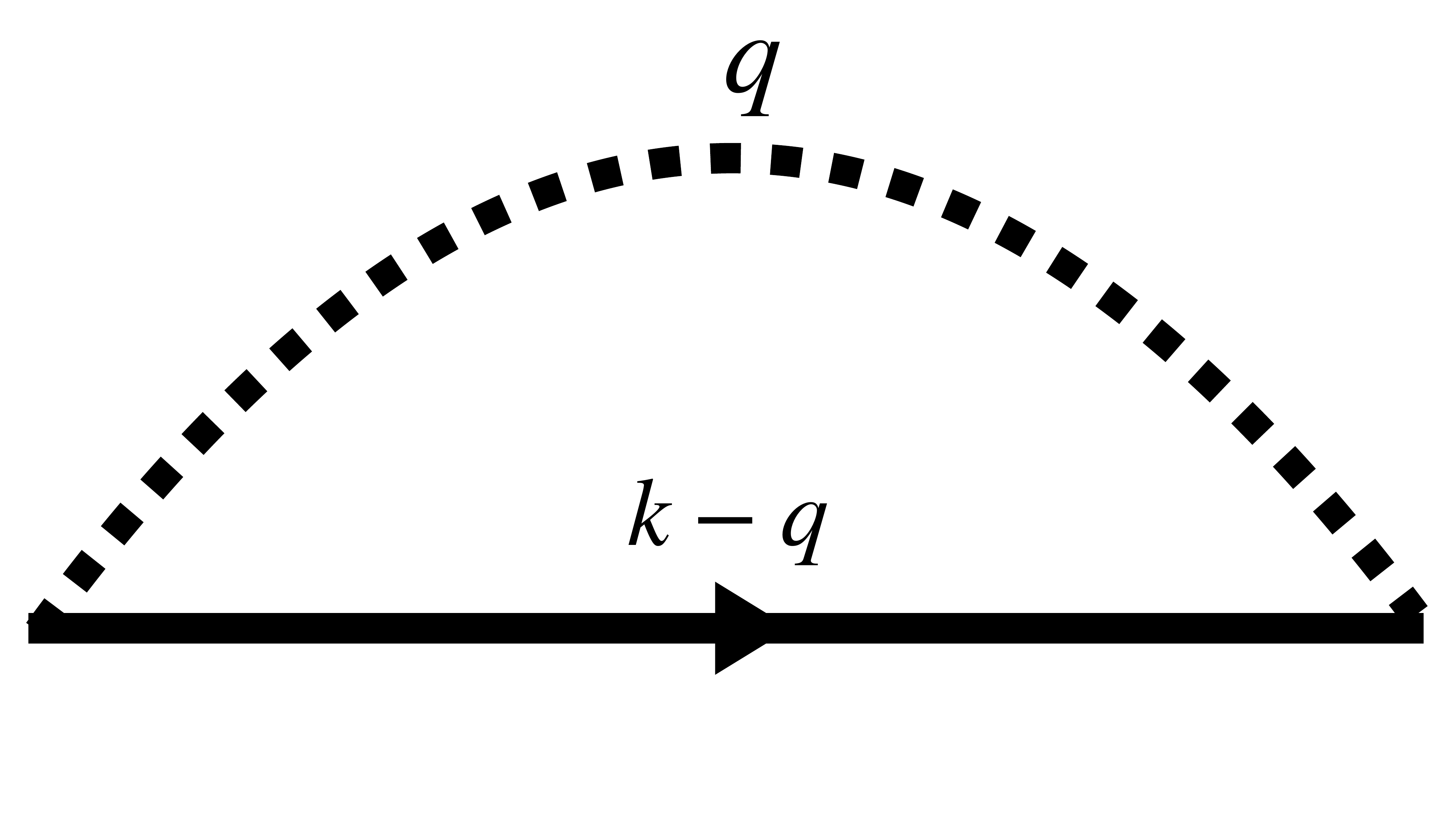}
    \includegraphics[width=0.3\linewidth]{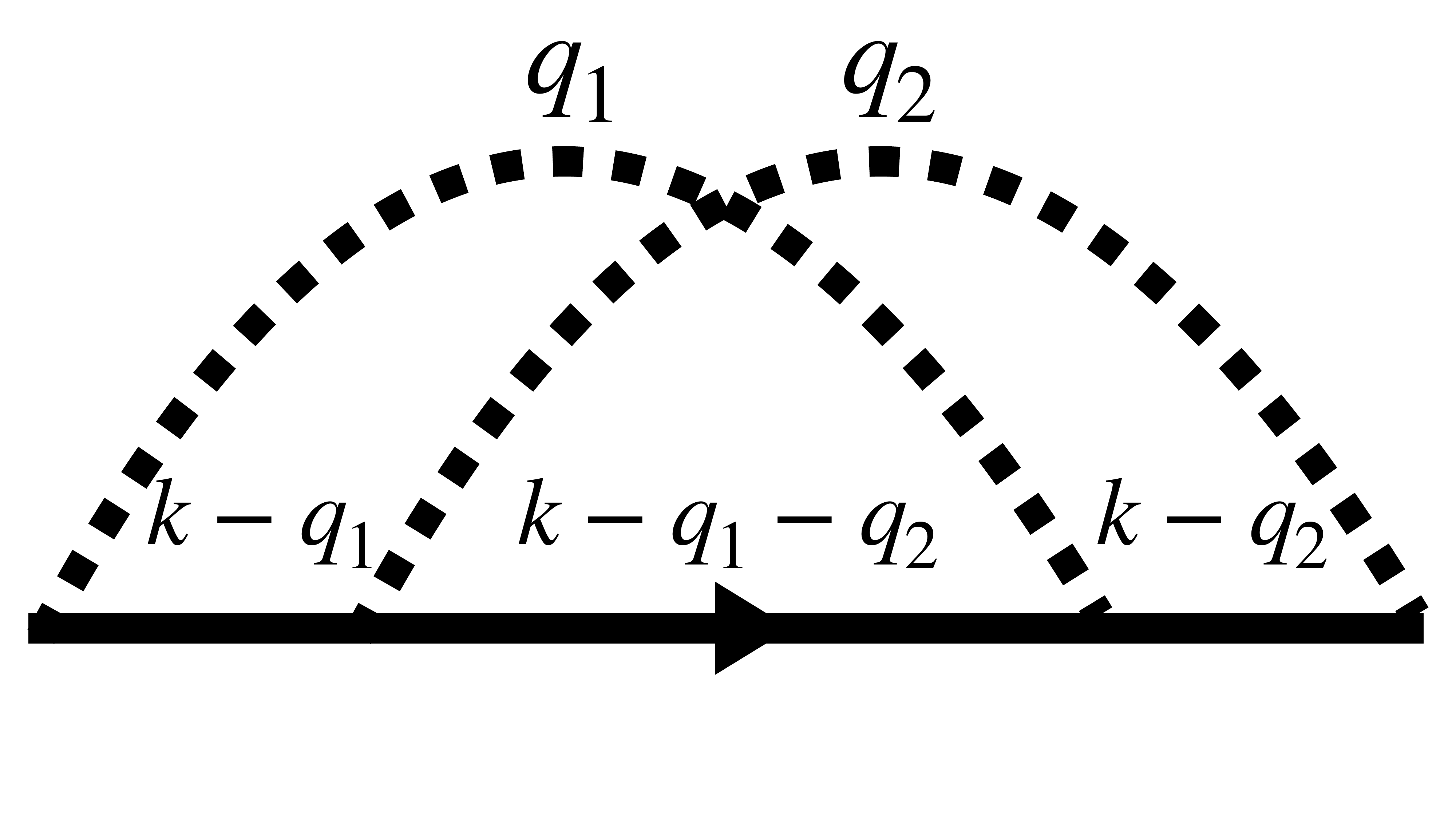}
    \includegraphics[width=0.3\linewidth]{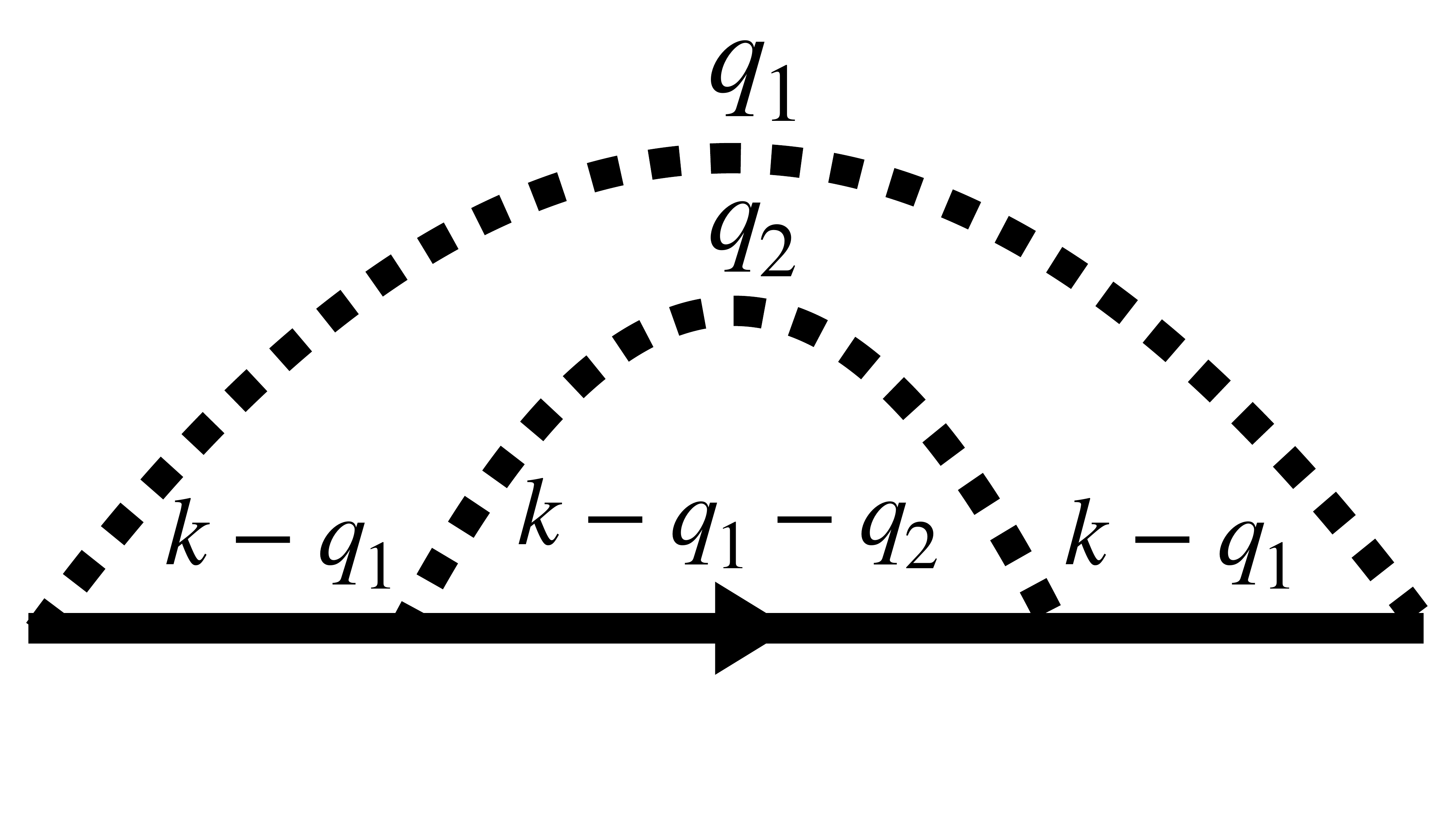}
    \caption{Self-energy $\Sigma(k,E)$ contribution up to second order in $S(q)$: (left) first-order contribution $\Sigma^{(1)}(k,E)$, (center) second order contribution with crossing disorder lines $\Sigma^{(2)}_{\mathrm{cross}}(k,E)$, (right) second order contribution without crossing disorder lines $\Sigma^{(2)}_{\mathrm{non-cross}}(k,E)$.}
    \label{fig:selfenergy}
\end{figure}

The individual contributions to the self-energy up to second order in $W^2 S(q)$ illustrated diagrammatically in Fig.~\ref{fig:selfenergy} are:
\begin{equation}
\label{eq:Sigma1_def}
    \Sigma^{(1)}(k,E) = W^2 \int_{-\pi}^{\pi} \frac{dq}{2\pi} S(q)G_0(k-q,E),
\end{equation}
\begin{multline}
    \Sigma^{(2)}_{\mathrm{cross}}(k,E) = W^4 \int_{-\pi}^{\pi} \frac{dq_1}{2\pi}\int_{-\pi}^{\pi} \frac{dq_2}{2\pi} S(q_1)S(q_2)\times\\
    G_0(k-q_1,E)G_0(k-q_2,E)G_0(k-q_1-q_2,E),
\end{multline}
\begin{multline}
    \Sigma^{(2)}_{\mathrm{non-cross}}(k,E) = W^4 \int_{-\pi}^{\pi} \frac{dq_1}{2\pi}\int_{-\pi}^{\pi} \frac{dq_2}{2\pi} S(q_1)S(q_2)\times\\
    [G_0(k-q_1,E)]^2G_0(k-q_1-q_2,E).
\end{multline}

According to Eq.~\eqref{eq:survtime}, we need to take the imaginary part of the self-energy and evaluate it on-shell, i.e., set $E = \epsilon_k$. This implies we have to consider the real and imaginary contributions to the bare propagator $G_0$,
\begin{equation}
    G_0(k', \epsilon_k) = \mathcal{P}\left(\frac{1}{\epsilon_k-\epsilon_{k'}}\right) -i\pi \delta(\epsilon_k-\epsilon_{k'}),
\end{equation}
where $\mathcal{P}$ denotes the principal value.
By virtue of the dispersion relation $\epsilon_k = -2 \cos k$, the imaginary part enforces having either no scattering ($k=k'$) or back-scattering $k=-k'$. This is a consequence of the 1D geometry, as there are no other angles available for scattering other than $0$ and $\pi$. 

For concreteness, we specify a particular form for the stealthy potential, which we choose to be stealthy hyperuniform, motivated by the connection we discussed earlier with optical transparency in layered media~\cite{klatt2025transparencyversusandersonlocalization}. We thus define, restricting to the first Brillouin zone,
\begin{equation}
\label{eq:stealthiness}
    S(q) = \Theta(|q|-k_0), \quad |q| < \pi,
\end{equation}
where the stealthy interval size $k_0$ is related to the $\chi$ parameter commonly used in stealthy hyperuniformity via $\chi = k_0/(2\pi)$. Notice that for $\chi = 0$, $w_q$ are independently distributed Gaussian random variables, and so are the $w_i$, recovering the totally uncorrelated case. For $\chi=1/2$, $w_i = (-1)^i$, corresponding to no disorder.
The spectrum of the Hamiltonian $H$ with $W=0$ is symmetric and double degenerate; therefore, we can focus on the interval $0 \le k <\pi/2$ without loss of generality.

We can now analyze in more detail the computation of the imaginary part of the self-energy $\mathrm{Im} \, \Sigma(k,\epsilon_k)$.
From Eq.~\eqref{eq:Sigma1_def}, we have
\begin{align}
    \mathrm{Im} \Sigma^{(1)}(k,\epsilon_k) &= W^2 \int_{-\pi}^{\pi} \frac{dq}{2 \pi} S(q) \ \mathrm{Im} G_0(k-q,\epsilon_k)\\
    &= -\pi W^2 \int_{-\pi}^{\pi} \frac{dq}{2 \pi} S(q)\delta(\epsilon_k-\epsilon_{k-q})
\end{align}
As discussed before, the dispersion relation $\epsilon_k = -2 \cos k$ imposes $q=0 \; \mathrm{mod} \,2\pi$, corresponding to no scattering, or $q=2k \; \mathrm{mod} \,2\pi$, meaning back-scattering, which is responsible for localization. Therefore we get
\begin{equation}
    \mathrm{Im} \Sigma^{(1)}(k,\epsilon_k) = - \frac{W^2}{4 \sin k} (S(0) + S(2k)), 
\end{equation}
and the inverse mean survival time, through second order in $W$, is given by
\begin{equation}
    \frac{1}{\tau(k)} = \frac{W^2}{2 \sin k} S(2k),
\end{equation}
where we dropped the forward scattering contribution $S(0)$, which does not contribute to the localization length. We then finally get
\begin{equation}
\label{eq:first-order}
    \frac{1}{\xi(k)} = \frac{W^2}{8 \sin^2 k} S(2k).
\end{equation}
A few comments are in order. 
The result in Eq.~\eqref{eq:first-order} is well known, and coincides with the finding of Izrailev and Krokhin~\cite{izrailev_localization_1999}. This result can also be derived from Boltzmann conductivity~\cite{herbut2000commentlocalizationmobilityedge}, and using Fermi's Golden Rule~\cite{Licciardello_1975_Conductivity,scardicchio2017perturbationtheoryapproachesanderson}. Note also that the expression in Eq.~\eqref{eq:first-order} is analogous to the one found within the strong-contrast expansion for the dynamic dielectric constant in 1D stealthy hyperuniform two-phase media~\cite{kim_effective_2023}, where the power spectrum at momentum $2k$ is replaced by the spectral density at momentum $2k$.

In the case of stealthy hyperuniform disorder we are interested in, it is possible to have no direct back-scattering at small $k$, as $S(q) = 0$ for $q < k_0 = 2\pi \chi$. Therefore, if $k_0>2k$, the right-hand side of Eq.~\eqref{eq:first-order} vanishes, and one needs to consider higher-order contributions to the self-energy. As a sanity check, notice that the middle of the spectrum corresponds to $k=\pi/2$, and one can avoid single back-scattering at $E \simeq 0$ only if $\chi \simeq 1/2$.

Let us then consider the case $k_0>2k$ and study the second-order correction to the self-energy, $\Sigma^{(2)}(k,\epsilon_k)$, and consider the diagram $\Sigma^{(2)}_{\mathrm{cross}}(k,\epsilon_k)$---the discussion for $\Sigma^{(2)}_{\mathrm{non-cross}}(k,\epsilon_k)$ is analogous. 
Since we are interested in the imaginary part of $\Sigma^{(2)}_{\mathrm{cross}}(k,\epsilon_k)$, there are two possibilities: either we take the imaginary parts of all three Green's functions, or the imaginary part of one and the real part of the other two. 
The former imposes $\delta(\epsilon_k-\epsilon_{k-q_1})\delta(\epsilon_k-\epsilon_{k-q_2})\delta(\epsilon_k-\epsilon_{k-q_1-q_2})$, which can be satisfied only if $q_1=q_2=0$---so no back-scattering---or if $q_1 = 2k$ and $q_2 = 0$---which then gives $S(2k) = 0$ since $2k<k_0$. 
We can instead obtain a non-trivial result when considering $\mathrm{Im} \, G_0(k-q_1-q_2,E)$ and the real part of $G_0(k-q_1,E)$ and $G_0(k-q_2,E)$. This gives
\begin{multline}
     \mathrm{Im}\, \Sigma^{(2)}_{\mathrm{cross}}(k,\epsilon_k) = -\pi W^4 \int_{-\pi}^{\pi} \frac{dq_1}{2\pi}\int_{-\pi}^{\pi} \frac{dq_2}{2\pi} S(q_1)S(q_2)\times\\
    \mathcal{P}\left(\frac{1}{\epsilon_k - \epsilon_{k-q_1}}\right)\mathcal{P}\left(\frac{1}{\epsilon_k - \epsilon_{k-q_2}}\right)\delta(\epsilon_k-\epsilon_{k-q_1-q_2}).
\end{multline}
Therefore, one can now get back-scattering by satisfying the equation
\begin{gather}
    q_1+q_2 = 2k\ \mathrm{mod}\, 2\pi, \\ 
    {\rm where} \; k_0 \le | q_{1,2}| < \pi \; {\rm and} \; 2k< k_0. \nonumber
\end{gather}
The first constraint comes from the stealthy hyperuniform disorder, and the second constraint is given by the perturbation theory order. Clearly, there is a solution only for $q_1>0$ and $q_2 <0$ or vice versa. Such an equation can be satisfied only if $k_0 < \pi-k$. It is easy to show that $\Sigma^{(2)}_{\mathrm{non-cross}}(k,\epsilon_k)$ gives the same conditions. Henceforth, one can compute explicitly the correction to $\xi(k)$ at second order, which will be of the form $\xi \sim W^{-4}$. We report the calculation in the Supplemental Material.

We can see, however, that not all possible combinations of $k$ (or energy $\epsilon_k$) and $k_0$ (or $\chi$) are included up to second order. This means that there are values of energy and size of the stealthiness parameter $\chi$ for which the first non-zero contribution to the $\xi$ occurs at third or higher order, corresponding to $\xi \sim W^{-2n}$ for $n\ge 3$. For instance, assume that $k_0 > \pi - k$ (given by $\Sigma^{(2)}=0$) and, at the same time, $k_0 > 2k$ (given by $\Sigma^{(1)}=0$): then computing $\Sigma^{(3)}(k,\epsilon_k)$, one finds that, in this range of parameters, it is possible to have back-scattering if there is a solution to the equation
\begin{gather}
\label{eq:rangeEq}
    q_1 + q_2 + q_3 = 2k\ \mathrm{mod}\, 2\pi, 
    \\ {\rm where} \; k_0 \leq |q_{1,2,3}| < \pi \; {\rm and} \; \pi-k_0 < k < \frac{k_0}{2}.  \nonumber  
\end{gather}
Equation~(\ref{eq:rangeEq}) can be satisfied either for $(3k_0 - 2\pi)/2 < k< k_0/2$ if $4\pi/5 < k_0 < \pi$ or for $\pi - k_0 < k < k_0/2$ if $2\pi/3 < k_0 < 4\pi/5$. For this range of parameters, $\xi \sim W^{-6}$.

Once again, not all possible values of $k$ and $k_0$ give a non-vanishing contribution to $\mathrm{Im}\, \Sigma(k,\epsilon_k)$ up to third order. To find the leading non-zero contribution to $\xi$, higher orders need to be included. 
By studying the equations corresponding to Eq.~\eqref{eq:rangeEq} up to fifth-order in perturbation theory, we can infer that the boundaries between different regions are
\begin{equation}
\label{eq:separatixes}
    k=
    \begin{cases}
        \frac{a}{2}(k_0-\pi) + \frac{\pi}{2}, \quad a = n\pi, \quad n \in \mathbb{N},\\
        b(\pi - k_0), \quad b = m\pi, \quad m \in \mathbb{N}.
    \end{cases}
\end{equation}

We can represent our findings in a diagram, showing, for each value of $k$ (or energy) and $k_0$ (or stealthy parameter $\chi$), the scaling of the localization length $\xi$ to leading order in $W$, as shown in Fig.~\ref{fig:phasediagr}. We present the diagram for other types of stealthy disordered potentials in the Supplemental Material.
\begin{figure}
    \centering
    \includegraphics[width=\linewidth]{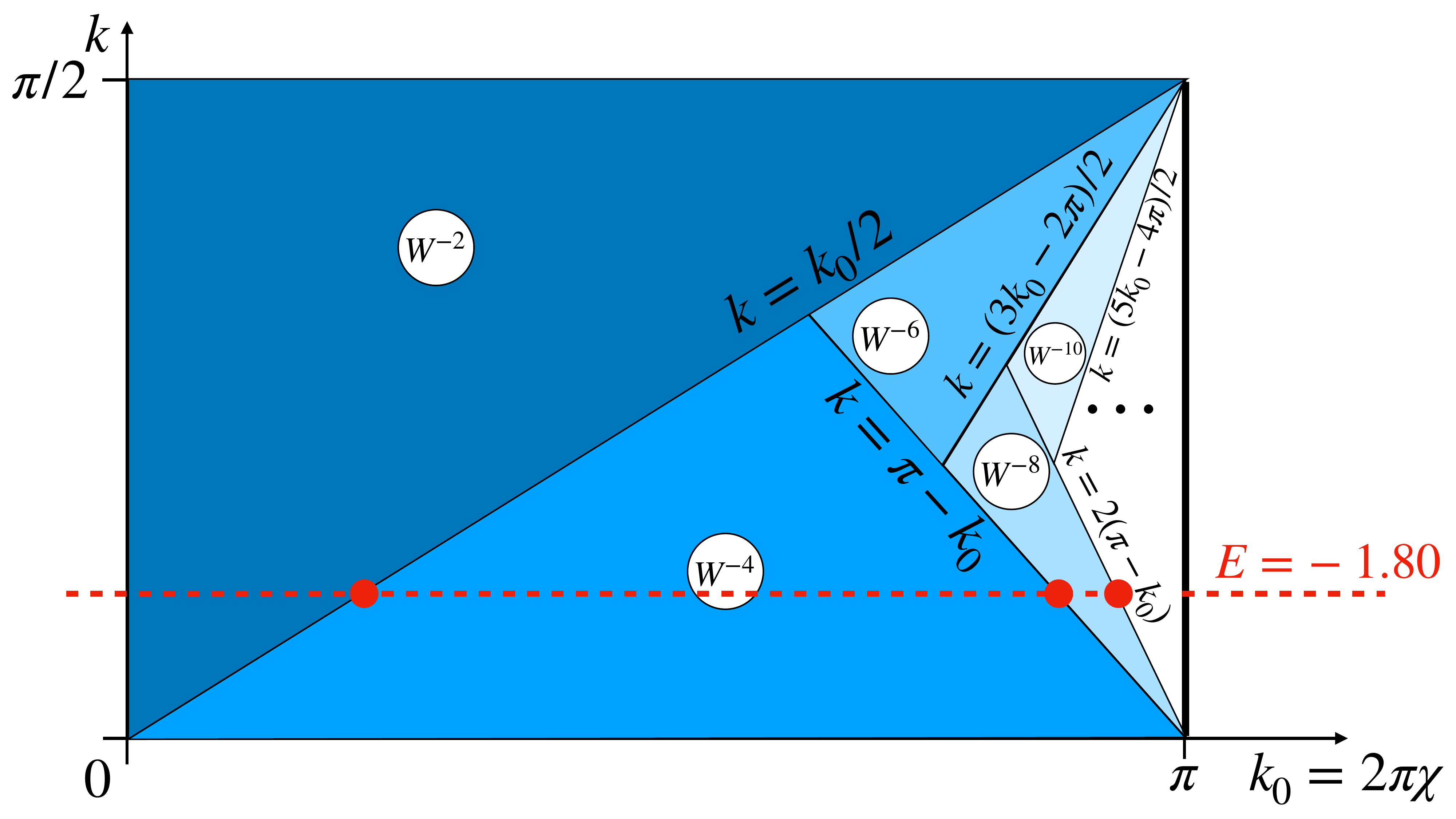}
    \caption{Phase diagram showing the dependence of the localization length $\xi(k,k_0)$ on the disorder strength $W$ at the leading non-vanishing order in perturbation theory. We also indicate the expression for the separatrices between different regions, whose general expression is given in Eq.~\eqref{eq:separatixes}. The red dotted line indicates the energy for which we will present numerical results.}
    \label{fig:phasediagr}
\end{figure}
Different regions represent different leading order non-zero contributions to the localization length, indicated by the power in the white circles. The scaling with $W^{-2}$ is the usual Anderson localization behavior that one also gets with ordinary (non-stealthy) disorder. The vertical line at $k_0 = \chi = 0$, corresponding to uncorrelated disorder, always exhibits this behavior, as expected.

The inverse localization length can be expressed as an asymptotic series in powers of $W^{2}$ as
\begin{equation}
\label{eq:seriesxi}
    \frac{1}{\xi(k,k_0)} \sim \sum_{n=1}^{\infty} a_{2n}(k,k_0) \,W^{2n}.
\end{equation}
Along the line $k_0 = \chi = 0$, each term in the series expansion of Eq.~\eqref{eq:seriesxi} is non-vanishing.
By increasing the value of $\chi$, one gets different scalings of the localization length depending on the energy. This means that, for those values of $k$ and $k_0$, some of the lower order terms $a_{2n}(k,k_0)$ in Eq.~\eqref{eq:seriesxi} are vanishing. 
In approaching the line $k_0=\pi$, i.e. $\chi = 1/2$, which corresponds to the lattice (no disorder), the index $m$ such that $a_{2n} = 0$ for $n<m$ is increasing, and ultimately $m \to \infty$ for $\chi \to 1/2$.

The series in Eq.~\eqref{eq:seriesxi} is an asymptotic expansion in small $W^2$, which is diverging when non-perturbative effects are excluded, as very often occurs in perturbative calculations~\cite{Dyson1952Divergence}. Once such non-perturbative effects are taken into account, the divergence is removed and the result can be written as a finite number of terms up to $n=N$ plus a small remainder, which is of order $o(W^{2N})$. When doing so, the vanishing of the lower order terms due to the stealthiness condition implies a larger localization length, scaling with the first non-zero term in $W^2$. In the next Section, we will show that this is indeed the case by numerically computing the dependence of the localization length $\xi$ on $W$ at different values of $\chi$, which properly accounts for the remainder terms.

In this section, we have discussed the series of transitions in the leading order contribution to the localization length at small disorder $W$.  In the Supplementary Material, we show that analogous transitions occur in the level splitting of adjacent energy levels that become degenerate as $W \to 0$ due to the $k\to -k$ symmetry. 

\section{Numerical results}

An eigenstate $\psi_{\alpha}$ of the Hamiltonian $H$ with eigenvalue $E_{\alpha}$ is exponentially localized with localization length $\xi_{\alpha}$ if $|\psi_{\alpha}|^2 \sim e^{r/\xi_{\alpha}}/\xi_{\alpha}$. The participation entropy is defined as $S_{\alpha} = -\int_0^{\infty} dr |\psi_{\alpha}(r)|^2 \ln |\psi_{\alpha}(r)|^2$, and we can define the corresponding fractal dimension 
\begin{equation}
\label{eq:fracdim}
    D_{\alpha} \equiv \frac{S_{\alpha}}{\ln L} \simeq \frac{1 + \ln \xi_{\alpha}}{\ln L},
\end{equation}
which corresponds to the usual definition of fractal dimension $D_1$ found in the literature~\cite{Chalker1996Spectral,Bogomolny2011Eigenfunction,kutlin2024investigating}. 

We compute numerically the fractal dimension from the eigenstates of the Hamiltonian and extract the localization length from the slope of the fractal dimension as a function of $1/\ln L$ for large $L$, according to Eq.~\eqref{eq:fracdim}. We consider system sizes up to $L=800000$ and average the obtained $D_{\alpha}$ over at least $1000$ disorder realizations and over $20$ eigenstates around the target energy $E_{\alpha}$.

\begin{figure}
    \centering
    \includegraphics[height=0.7\linewidth]{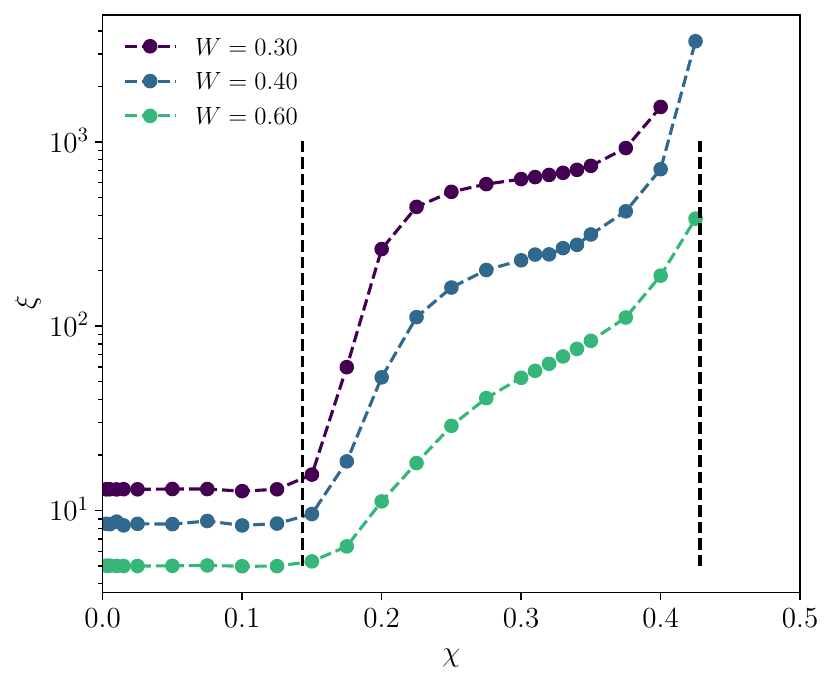}
    \caption{Localization length as a function of $\chi$ for different values of $W$. The vertical lines are positioned where the perturbation theory predicts a change in the first non-vanishing contribution (corresponding to the first two red dots from the left in Fig.~\ref{fig:phasediagr}). Further transitions require system sizes larger than the numerically accessible one.}
    \label{fig:loclength_chi}
\end{figure}
\begin{figure}
    \centering
    \includegraphics[height=0.7\linewidth]{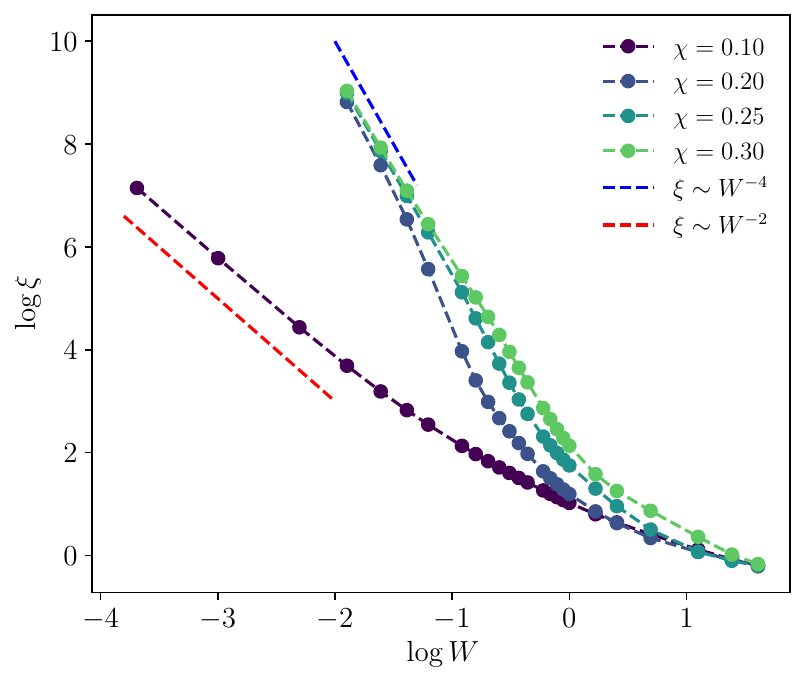}
    \caption{Localization length as a function of $\log W$ for different values of $\chi$, as shown in the legend, obtained for eigenstates at energy $E = -1.8$, corresponding to the horizontal red line in Fig.~\ref{fig:phasediagr}. For $\chi=0.1$, the perturbation theory predicts that $\xi \sim W^{-2}$, as confirmed by the numerical result (see red dashed line, which is a guide for the eye). For $\chi=0.2, \, 0.25,\, 0.3$ the perturbation theory predicts $\xi \sim W^{-4}$, as confirmed by the numerical result.}
    \label{fig:loclength_W}
\end{figure}

We show in Fig.~\ref{fig:loclength_chi} the localization length as a function of $\chi$ for a few values of $W$, where the energy is fixed at $E=-1.8$, corresponding to the red dashed line in the top panel of Fig.~\ref{fig:phasediagr}. By increasing $\chi$, the localization length $\xi$ shows sudden increases. Notice that, by decreasing $W$, the transitions in numerical curves are becoming steeper and the $\chi$ values where they occur are going towards the perturbation theory prediction, reported as vertical black dashed lines, corresponding to the first two red dots from the left in the top panel of Fig.~\ref{fig:phasediagr}.
Subsequent transitions are not visible due to the limited system sizes that can be reached numerically.

In Fig.~\ref{fig:loclength_W}, we report the scaling of $\xi$ with $W$ for a few values of $\chi$. For $\chi = 0.1$, the perturbation theory predicts the scaling $\xi \sim W^{-2}$, as indeed confirmed from the numerical results. For higher values of $\chi = 0.2,\, 0.25,\, 0.3$, the perturbation theory predicts $\xi \sim W^{-4}$, as confirmed once again by the numerical data. Therefore, we have confirmed that perturbation theory correctly predicts the scaling $\xi \sim W^{-2n}$ and the transition from $n=1$ to $n=2$.

\section{Discussion}

Generically, disordered stealthy hyperuniform materials lack Bragg peaks similar to liquid-like disordered systems and yet suppress long-range density fluctuations as crystals.
Our results indicate that stealthy disorder represents an intermediate regime when it comes to localization in 1D quantum disordered systems as well. The stealthiness parameter $\chi$ acts as a control knob that varies between $\chi = 0$, corresponding to a random potential whose values are uncorrelated, and $\chi = 1/2$, corresponding to a periodic potential. 
Stealthy disorder leads to effective delocalization, in which the localization length scales as $W^{-2n}$ where $n>1$.

More precisely, for an arbitrarily large system size $L$, fixed small disorder $W$, and fixed energy $E$, there exists a stealthy interval size $\chi < 1/2$ for which $a_{2n} = 0$ (see Eq.~\eqref{eq:seriesxi}) for all $n<m$ with $m$ arbitrarily large, thus meaning that the localization length $\xi$ exceeds the system size $L$. Therefore, for all practical purposes, there is a finite continuous range of effectively extended states, associated with the energy $E$, and the system is effectively delocalized. In the limit $\chi \rightarrow 1/2$, we expect non-perturbative effects not captured by our treatment to make non-negligible contributions.  We leave the investigation of this region of the phase diagram for future work.

We obtained the aforementioned result by computing the localization length in the 1D Anderson model with stealthy disorder, such that its power spectrum vanishes for a continuous band.
From the perturbative expansion of the imaginary part of the self-energy, we were able to determine, for each value of the energy and of the stealthiness parameter $\chi$, the leading contribution $W^{-2n}$, $n\ge 1$, to the localization length $\xi$, summarized in Fig.~\ref{fig:phasediagr}.
We supported our findings with a numerical analysis of the localization length extracted from the fractal dimension of eigenstates of a given energy. 
Our results show that stealthy disorder makes it possible to tune how the localization length scales, enabling it to become larger than the system size even when the disorder variance is held fixed. Unlike uncorrelated disorder, for which $\xi$ scales as $W^{-2}$ to leading order for small $W$, the result for stealthy disordered systems is that the leading order terms in the perturbation expansion of $\xi$ vanish identically for a progressively large number of terms as $\chi$ increases such that $\xi$ scales as  $W^{-2n}$ with arbitrarily large $n$, where $n$ is a positive integer.

Our findings for the Anderson model are reminiscent of recent studies on the propagation of light in a one-dimensional layered medium~\cite{kim_effective_2023,torquato_nonlocal_2021,klatt2025transparencyversusandersonlocalization}. The disorder strength $W$ in the Anderson problem is analogous to the dielectric contrast in the two-phase layered medium~\cite{torquato_nonlocal_2021}. In the latter case, complete transparency can be shown to occur at least up to third order in the strong dielectric contrast expansion~\cite{kim_effective_2023,torquato_nonlocal_2021} for a finite continuous range of frequencies, which matches the result for the tight-binding Anderson model where contributions to the localization length are zero to $2n$-th order for arbitrarily large finite $n$. Moreover, the momentum dependence of the imaginary part of the self-energy in the Anderson problem parallels that of the nonlocal attenuation function in the 1D stealthy two-phase medium~\cite{kim_effective_2023}.     

From a practical perspective, the Anderson model with stealthy disorder can be realized on currently available experimental platforms, such as ultracold atoms in optical lattices, waveguide arrays, photonic lattices in photorefractive crystals in the paraxial regime of propagation, or other programmable quantum simulators, which provide a high degree of tunability of the potential~\cite{Billy2008,roati2008anderson, Sc07b, lahini_anderson_2008, Schreiber2015Observation,Smith2016}. In addition, stealthy hyperuniformity was recently realized and probed experimentally in a two-dimensional photonic crystal slab platform~\cite{barsukova2025stealthy}. 
We leave the intriguing question of stealthy hyperuniformity in interacting quantum systems for future work.

\acknowledgments

We are grateful to Federico Balducci for his useful suggestions and collaboration on related topics. We also acknowledge fruitful discussions with Miguel Gonçalves, David Huse, Jaeuk Kim, and Antonello Scardicchio.
The work of JK, MCR, PJS, ST, and CV was supported by the Army Research Office
under Cooperative Agreement No. W911NF-22-2-0103.

\bibliography{references.bib}
\pagebreak
\widetext
\newpage

\section*{--- Supplemental Material ---}

\subsection{Stealthy non-hyperuniform potentials}

In the main text, we discussed in detail the perturbation theory calculation of the localization length and the corresponding numerical results in the case of stealthy hyperuniform disorder, motivated by the recent results on optical properties in layered 1D media~\cite{klatt2025transparencyversusandersonlocalization}.
As stated in the main text, the method we presented is valid for any stealthy potential, i.e., with a continuous interval of wave numbers for which the power spectrum vanishes.

In this Section, we report the phase diagram, equivalent to Fig. 3 of the main text, for other types of stealthy potentials. 
\begin{figure}[h!]
    \centering
    \includegraphics[width=\linewidth]{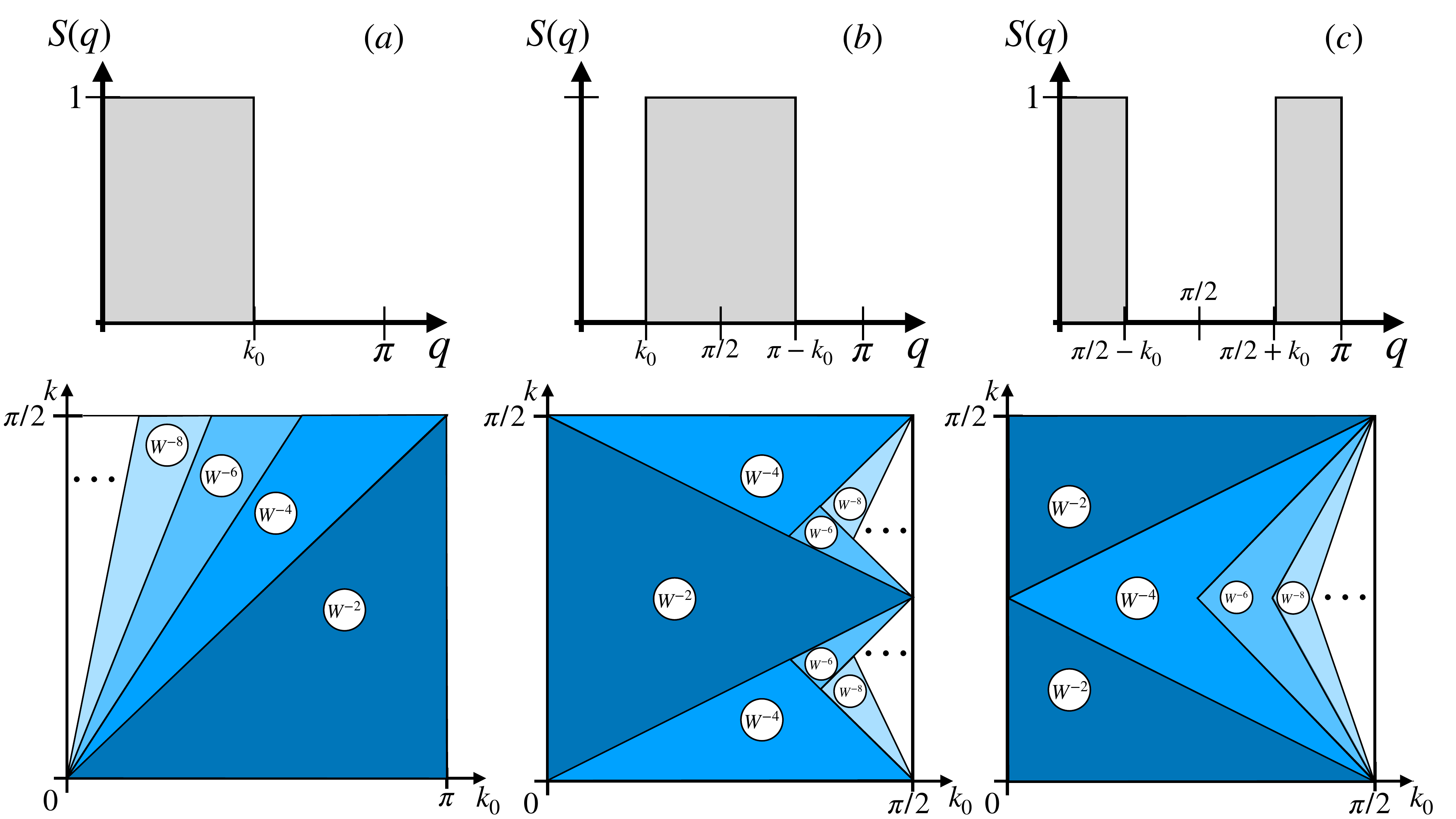}
    \caption{Different types of stealthy disordered correlations. }
    \label{fig:Kq_suppl}
\end{figure}
In Fig.~\ref{fig:Kq_suppl} we show three different types of disordered stealthy potentials, whose graphical representation is reported in the first row and labeled (a)-(c).
For each of them, we show the resulting diagram in the bottom row, obtained following the same procedure presented in the main text for the hyperuniform case.
Figure~\ref{fig:Kq_suppl} (a) represents a non-hyperuniform stealthy disorder, for which $S(q) = 0$ for $|q|>k_0$. The different position of the stealthy interval leads to a larger localization length near the middle of the spectrum $k=\pi/2$, rather than the edge $k=0$, as we showed in the main text. This result is equivalent to the one presented in Ref.~\cite{Lugan2009Onedimensional} for the case of speckle potentials.
Figure~\ref{fig:Kq_suppl} (b) shows a stealthy hyperuniform disorder, in which the power spectrum vanishes both near $q=0$ and $q=\pi$, and is often called “multi-stealthy". In this case, both the center and the edge of the spectrum display a larger localization length, as visible in the diagram. In Figure~\ref{fig:Kq_suppl} (c), we show another example of non-hyperuniform stealthy disorder, in which the power spectrum vanishes around $q=\pi/2$. In this case, both the edge and the middle of the spectrum maintain a small localization length $\xi \sim W^{-2}$, while for intermediate energies $k \sim \pi/4$ it becomes increasingly larger, $\xi \sim W^{-2n}$, $n \ge 2$.

The results shown in Fig.~\ref{fig:Kq_suppl} confirm that the stealthiness condition is instrumental for the suppression of back-scattering, even in the non-hyperuniform case. Also in these cases, it is possible to find a range of values of $k_0$ for which the potential is disordered and the localization length can be made larger than the system size for a fixed energy and disorder strength.

\subsection{Details of the numerical phase diagram}
\begin{figure}[h!]
    \centering
    \includegraphics[width=.48\linewidth]{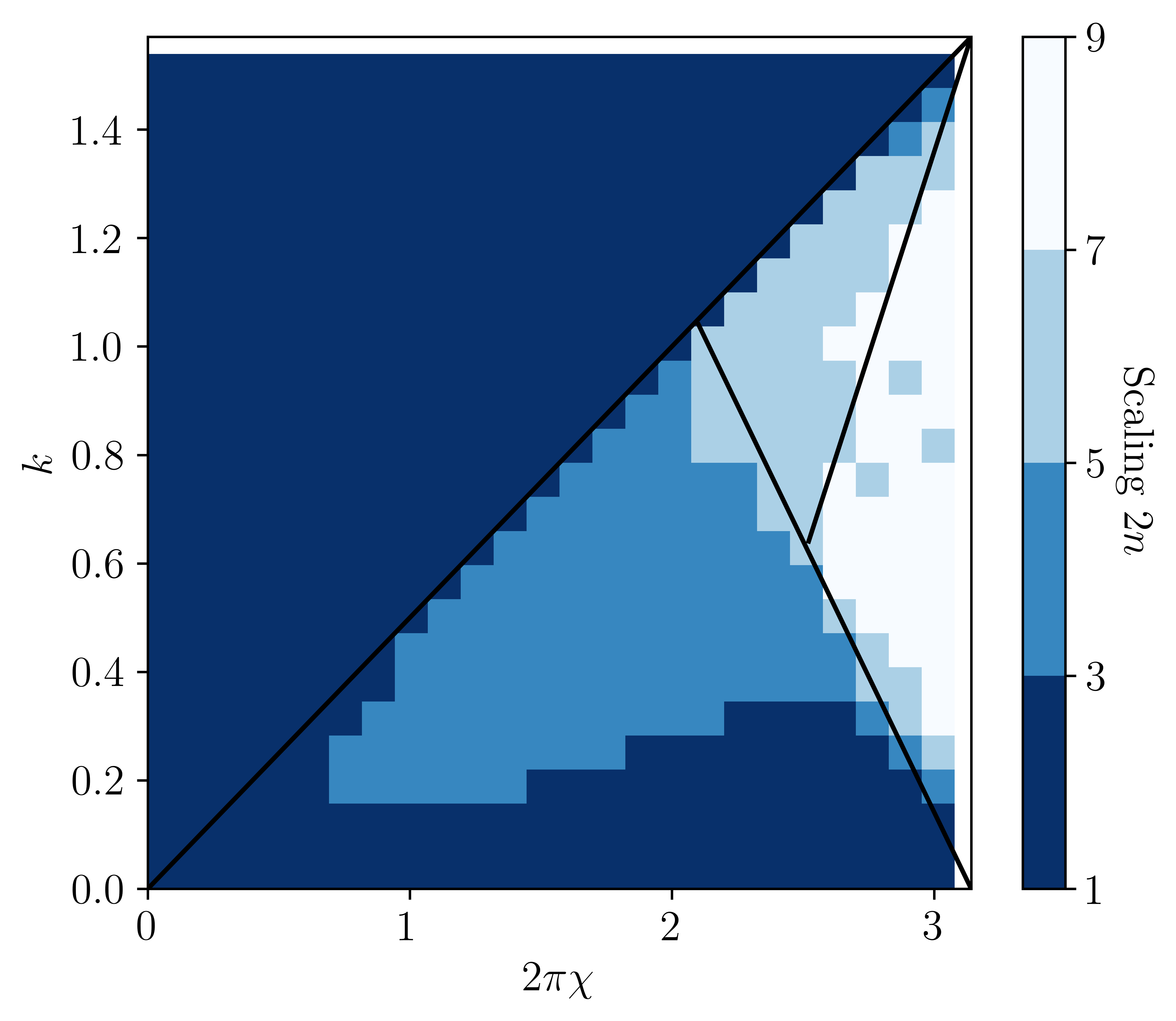}
    \includegraphics[width=.48\linewidth]{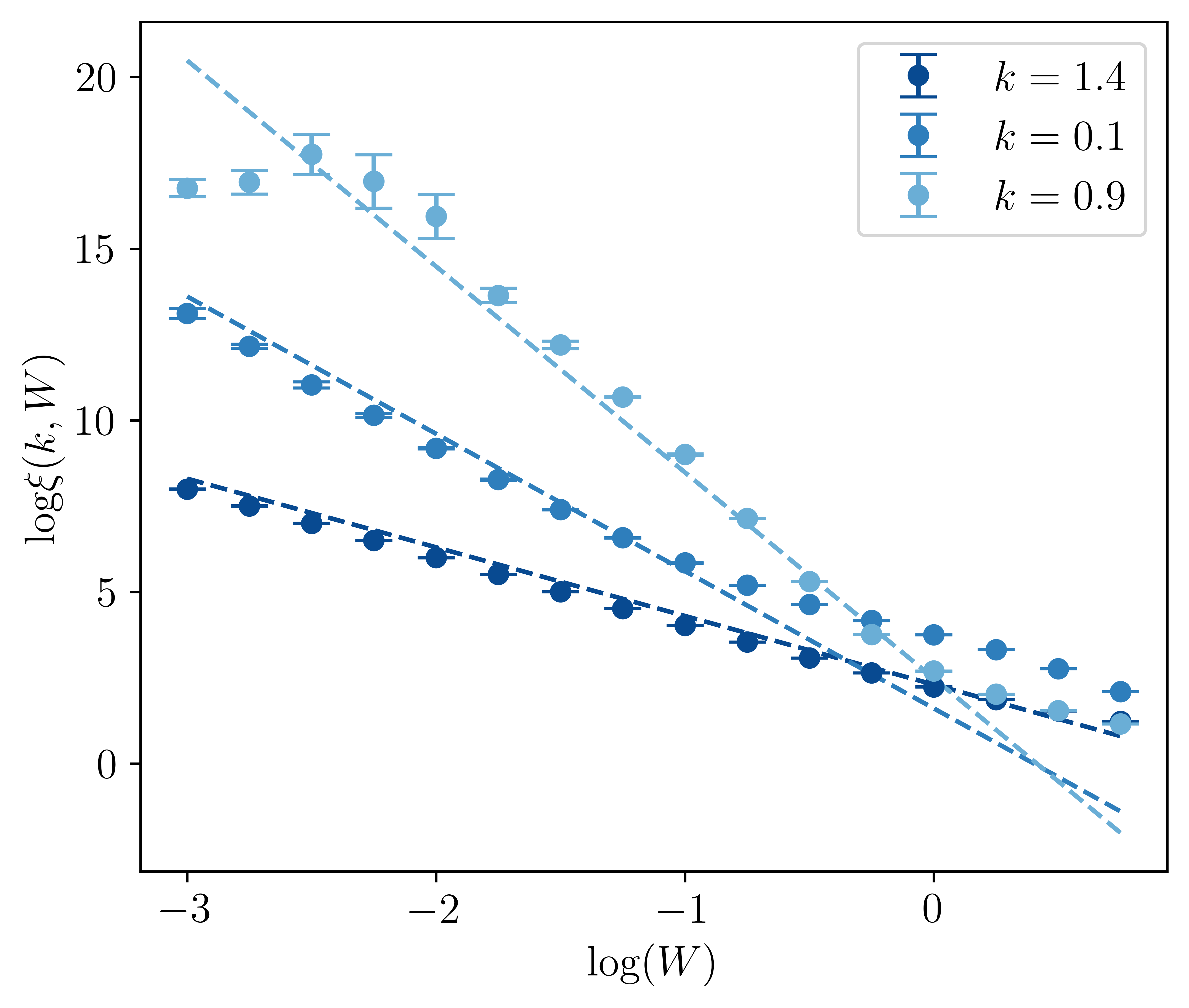}
    \caption{{\it(Left)} Numerical phase diagram, obtained by computing the localization length with the transfer matrix method. More details are provided in the Supplemental Material. Near $k=0$, the fitting deviates from the theoretical prediction because the asymptotic behavior sets in beyond the sample size, $L=5\cdot 10^6$. In the gray region at large $k$, the localization length exceeds the numerical finite size. {\it(Right)} Data used for fitting the scaling exponent $n$ in the phase diagram in Fig.~\ref{fig:phasediagr} (bottom panel) for $\chi=0.33$ and for three values of $k$ in phases with $n=1,2,3$.}
    \label{fig:fits_suppl}
\end{figure}

In Fig.~\ref{fig:fits_suppl} (left), we show the phase diagram obtained by numerically computing the localization length with the transfer matrix method, where the scaling exponent $n$ (see Eq.~\eqref{eq:seriesxi}) is determined with a linear fit $\log \xi = - 2n \log W + C$ sweeping over the phase diagram in $25 \times 25$ resolution.
We obtained this full $(k,\chi)$ phase diagram by automatically fitting the localization length and extracting the value of $n$ for each $(k, \chi)$ possible. To capture phases with higher $n$, we choose a larger $L=5\cdot 10^6$ here. To achieve the larger system size efficiently, the transfer matrix method is used.
The discrepancy between the theoretical prediction and the results shown in Fig.~\ref{fig:fits_suppl}  in the region near $k=0$ is a numerical artifact due to the asymptotic scaling behavior requiring a larger sample size to set in for this range of $k$.

In Fig~\ref{fig:fits_suppl} (right), we show the data used for the fits at fixed $\chi=0.25$ and three values of $k$ in distinct phases with $n=1,2,3$. Up to $16$ data points in $W$ enter the fit; points with a localization length larger than the system are excluded from the fit.

\subsection{Disorder dependence in the level splitting}

As anticipated in the main text, the transitions observed in the leading contribution to the localization length in the perturbative expansion at small $W$ in the presence of stealthy disorder can also be observed in the energy levels of the model.
In the absence of disorder, i.e., $W=0$, the spectrum is doubly degenerate, due to the symmetry $k \to -k$, as $\epsilon_k = - 2 \cos k$.

Let us consider two degenerate levels $E_1^0 = E_2^0 = E^0$ at zero disorder, and let us denote with $\ket{\psi_1^0}$ and $\ket{\psi_2^0}$ the corresponding eigenstates. When a small disorder is turned on, the degeneracy is lifted (since there are no other symmetries). The splitting can be computed using degenerate perturbation theory, and, at first order, the new energies of $\ket{\psi_1}$ and $\ket{\psi_2}$ are
\begin{equation}
    E_1 = E_1^0 + |\bra{\psi_1} V \ket{\psi_2}|, \quad E_2 = E_2^0 -|\bra{\psi_1} V \ket{\psi_2}|
\end{equation}
so that
\begin{equation}
    \Delta E_{12} = 2|\bra{\psi_1} V \ket{\psi_2}|.
\end{equation}
In the case of uncorrelated disorder, the matrix element $\bra{\psi_1} V \ket{\psi_2} \neq 0$, and in particular is proportional to $W$, so that $\Delta E_{12} \propto W$. In the case of stealthy disorder, however, the matrix element $\bra{\psi_1} V \ket{\psi_2}$ can be zero if the potential $V$ has no Fourier components connecting $k$ to $-k$, namely if $\tilde{V}(2k) = 0$. When this happens, one needs to go to the second order in degenerate perturbation theory, and the effective potential connecting the two levels is
\begin{equation}
    V_{12}^{(\mathrm{eff})} = \sum_{q \neq 1,2} \frac{\bra{\psi_1} V \ket{q}\bra{q} V \ket{\psi_2}}{E^0 - E_q},
\end{equation}
so that $\Delta E_{12} \simeq 2 |V_{12}^{(\mathrm{eff})}| \propto W^2$. If even $V_{12}^{(\mathrm{eff})} = 0$ at second order due to the stealthiness condition, the third order should be considered, similar to the case of the imaginary part of the self-energy.

Note that $\Delta E_{12} \sim \sqrt{\frac{1}{\xi(E)}}$; {\it e.g.,}  when $\Delta E_{12} \propto W$, $1/\xi(E) \propto W^2$. This can be understood from Fermi's Golden Rule, stating that the transition rate between the states $\ket{\psi_1}$ and $\ket{\psi_2}$ with density of states $\rho(E^0)$ is given by
\begin{equation}
    \Gamma_{1\to 2} = 2\pi |\bra{\psi_1} V \ket{\psi_2}|^2 \rho(E^0).
\end{equation}
The mean survival time associated with backscattering is $\tau_{12} \sim 1/\Gamma_{1\to 2}$, and therefore the localization length scales as the inverse of $\Gamma_{1\to 2}$. When $\Delta E_{12} \propto W$, $\Gamma_{1\to2} \sim 1/\xi(E) \propto W^2$, etc.

\begin{figure}
    \centering
    \includegraphics[width=0.49\linewidth]{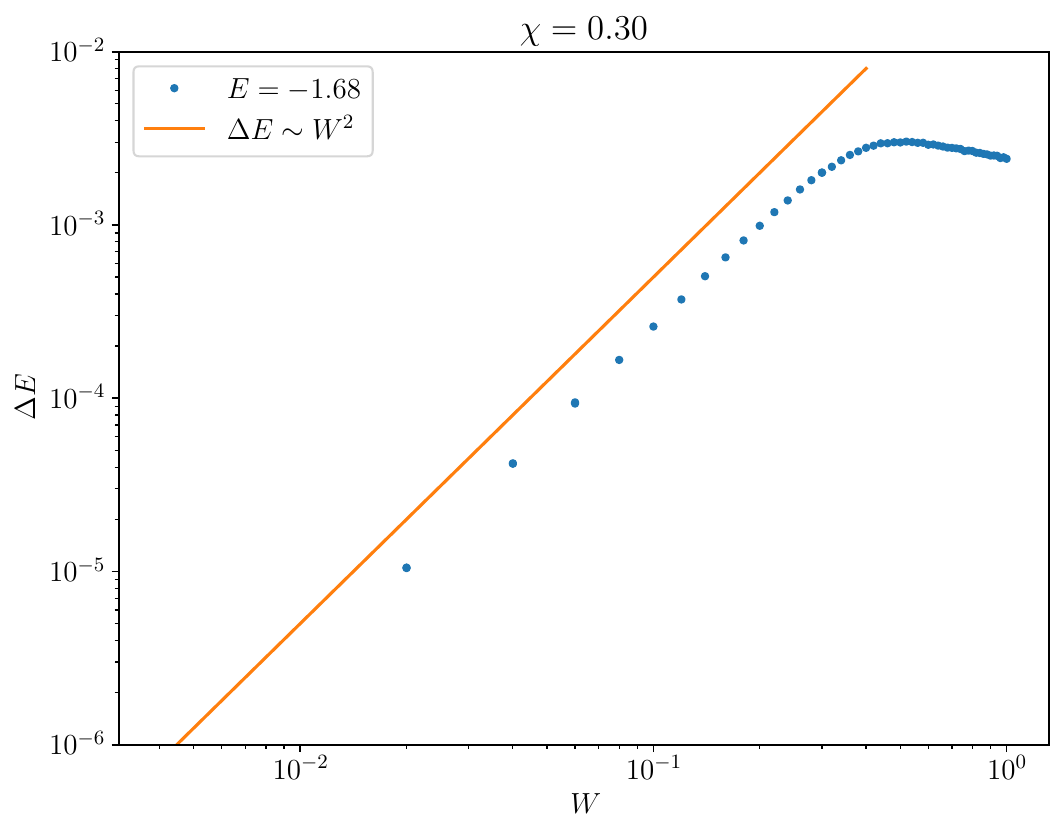}
    \includegraphics[width=0.49\linewidth]{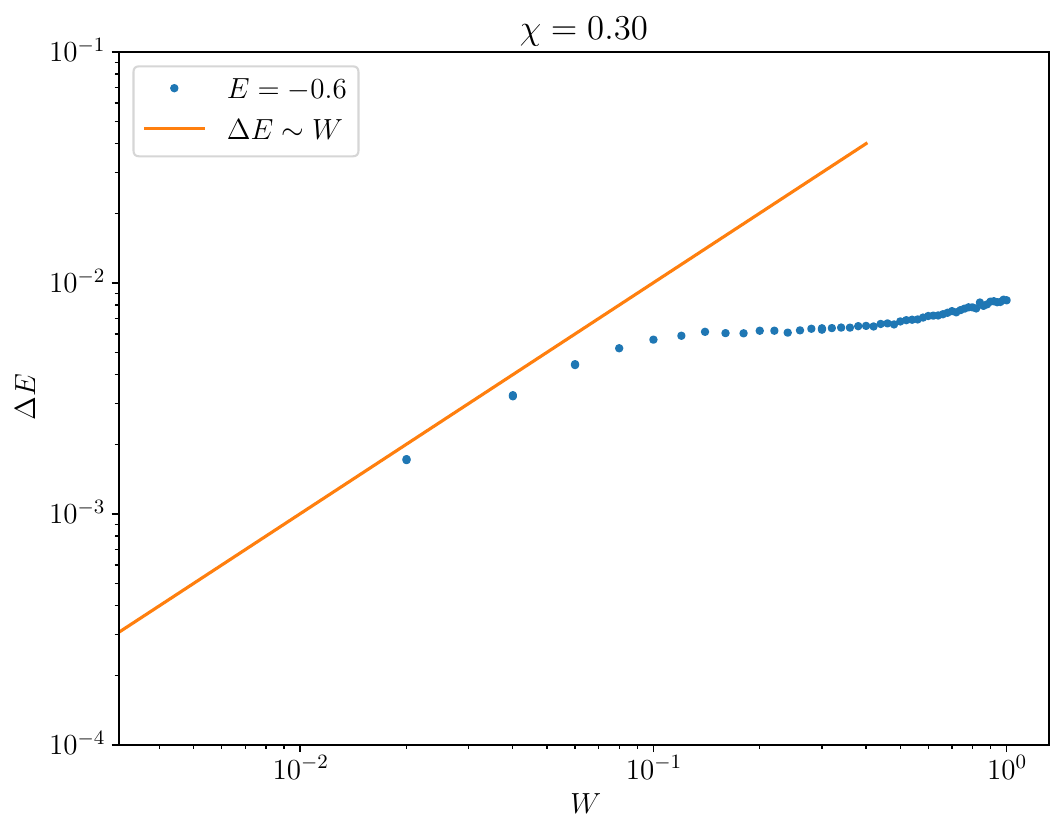}
    \caption{Level splitting of two degenerate energy levels in the zero disorder limit as a function of $W$. (Left) For $\chi = 0.3$ and small energy, such that $\xi \sim W^{-4}$, the level spacing is $\Delta E \sim W^2$. (Right) For the same stealthiness but bigger energy, such that $\xi \sim W^{-2}$, the level spacing $\Delta E \sim W$.}
    \label{fig:Ediff_W}
\end{figure}

\begin{figure}
    \centering
    \includegraphics[width=0.49\linewidth]{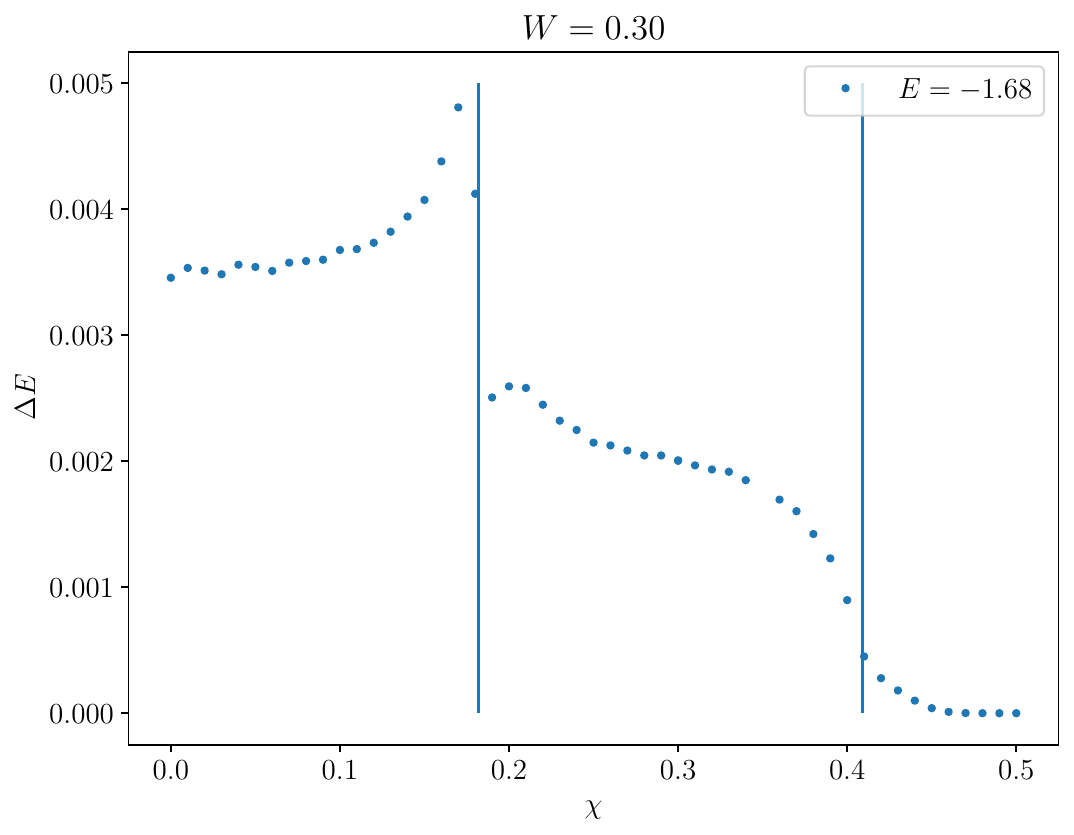}
    \includegraphics[width=0.49\linewidth]{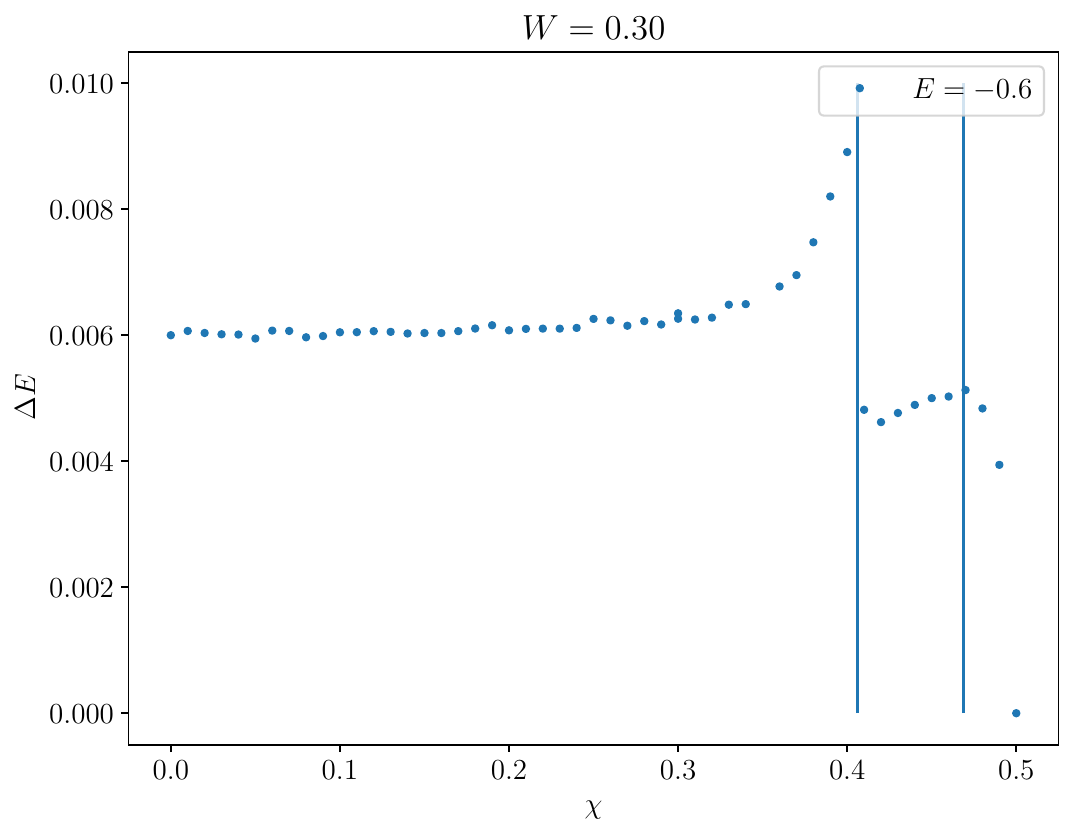}
    \caption{Level splitting of two degenerate energy levels in the zero disorder limit for fixed $W$ and across $\chi$. The vertical lines are determined by the perturbation theory for the self-energy, and the transitions observed in the level splitting coincide, both at small and large energies.}
    \label{fig:Ediff_chi}
\end{figure}

This can also be seen in the numerical data, as reported in Figs~\ref{fig:Ediff_W} and~\ref{fig:Ediff_chi}.

\subsection{Fourth order coefficient}

We report here the computation of the $W^4$ contribution to the inverse localization length, in the regime in which the $W^2$ contribution vanishes, namely $2k < k_0 < \pi - k$.
To do so, we need to compute the diagrams $\Sigma^{(2)}_{\mathrm{cross}}(k,E)$ and $\Sigma^{(2)}_{\mathrm{non-cross}}(k,E)$, given by
\begin{equation}
    \Sigma^{(2)}_{\mathrm{cross}}(k,E) = W^4 \int_{-\pi}^{\pi} \frac{dq_1}{2\pi}\int_{-\pi}^{\pi} \frac{dq_2}{2\pi} S(q_1)S(q_2)
    G_0(k-q_1,E)G_0(k-q_2,E)G_0(k-q_1-q_2,E),
\end{equation}
\begin{equation}
    \Sigma^{(2)}_{\mathrm{non-cross}}(k,E) = W^4 \int_{-\pi}^{\pi} \frac{dq_1}{2\pi}\int_{-\pi}^{\pi} \frac{dq_2}{2\pi} S(q_1)S(q_2)
    [G_0(k-q_1,E)]^2G_0(k-q_1-q_2,E).
\end{equation}
When $\Sigma^{(1)}(k,\epsilon_k) = 0$, we get, following our discussion in the main text,
\begin{equation}
    \mathrm{Im}\, \Sigma^{(2)}_{\mathrm{cross}}(k,\epsilon_k) = -\pi W^4 \int_{-\pi}^{\pi} \frac{dq_1}{2\pi}\int_{-\pi}^{\pi} \frac{dq_2}{2\pi} S(q_1)S(q_2)
    \mathcal{P}\left(\frac{1}{\epsilon_k - \epsilon_{k-q_1}}\right)\mathcal{P}\left(\frac{1}{\epsilon_k - \epsilon_{k-q_2}}\right)\delta(\epsilon_k-\epsilon_{k-q_1-q_2}),
\end{equation}
\begin{equation}
    \mathrm{Im}\, \Sigma^{(2)}_{\mathrm{non-cross}}(k,\epsilon_k) = -\pi W^4 \int_{-\pi}^{\pi} \frac{dq_1}{2\pi}\int_{-\pi}^{\pi} \frac{dq_2}{2\pi} S(q_1)S(q_2)
    \left[\mathcal{P}\left(\frac{1}{\epsilon_k - \epsilon_{k-q_1}}\right)\right]^2\delta(\epsilon_k-\epsilon_{k-q_1-q_2}).
\end{equation}
The $\delta$-function, in the case of back-scattering we are interested in, is given by
\begin{equation}
    \delta(\epsilon_k-\epsilon_{k-q_1-q_2}) = \frac{\delta(q_2-2k+q_1)}{2 \sin(k)},
\end{equation}
and thus
\begin{equation}
    \mathrm{Im}\, \Sigma^{(2)}_{\mathrm{cross}}(k,\epsilon_k) = -\frac{\pi W^4}{32\pi^2 \sin(k)} \int_{[-\pi,-k_0]\cup [k_0,\pi]} dq_1
    \left(\frac{1}{\cos(k) - \cos(k-q_1)}\right)\left(\frac{1}{\cos(k) - \cos(-k+q_1)}\right),
\end{equation}
\begin{equation}
    \mathrm{Im}\, \Sigma^{(2)}_{\mathrm{non-cross}}(k,\epsilon_k) = -\frac{\pi W^4}{32\pi^2 \sin(k)} \int_{[-\pi,-k_0]\cup [k_0,\pi]} dq_1
    \left(\frac{1}{\cos(k) - \cos(k-q_1)}\right)^2.
\end{equation}
The two integrals are equal, and thus
\begin{align}
    \mathrm{Im}\, \Sigma^{(2)}(k,\epsilon_k) &= -\frac{\pi W^4}{16\pi^2 \sin(k)} \int_{[-\pi,-k_0]\cup [k_0,\pi]} dq_1
    \left(\frac{1}{\cos(k) - \cos(k-q_1)}\right)^2\\
    &= -\frac{\pi W^4}{16\pi^2 \sin(k)} \left[  
    \frac{1}{2} \csc ^2(k) \left(-\csc (k) \sin \left(\frac{k_0}{2}\right) \left(\csc
   \left(k-\frac{k_0}{2}\right)+\csc \left(k+\frac{k_0}{2}\right)\right)\right.\right.\\
   & \qquad \qquad \left.\left.
   -2
   \cot (k) \log \left(\sin \left(k+\frac{k_0}{2}\right) \left(-\csc
   \left(k-\frac{k_0}{2}\right)\right)\right)+2 \cot
   \left(\frac{k_0}{2}\right)\right) \right].
\end{align}
The inverse localization length is then
\begin{align}
 \label{eq:second-order}
    \frac{1}{\xi(k)} = \frac{W^4}{32\pi \sin^2(k)} &\left[ 
    \frac{1}{2} \csc ^2(k) \left(-\csc (k) \sin \left(\frac{k_0}{2}\right) \left(\csc
   \left(k-\frac{k_0}{2}\right)+\csc \left(k+\frac{k_0}{2}\right)\right)\right.\right.\\
   &\left.\left.
   -2
   \cot (k) \log \left(\sin \left(k+\frac{k_0}{2}\right) \left(-\csc
   \left(k-\frac{k_0}{2}\right)\right)\right)+2 \cot
   \left(\frac{k_0}{2}\right)\right) \right].
\end{align}
One can compute the leading order term in $\pi - k_0$, or equivalently $1/2 - \chi$. By expanding the above expression around $k_0 = \pi$, we get.
\begin{align}
    \frac{1}{\xi(k)} = \frac{W^4}{32\pi \sin^2(k)}\frac{(\pi-k_0)}{2} \left[\sec ^2(k) - \frac{1}{24} (\pi - k_0 )^2 (\cos (2 k)-5)
   \sec ^4(k)+O\left((\pi -k_0)^3\right)\right]
\end{align}

\end{document}